\documentclass[12pt]{article}

\usepackage{amsmath}
\usepackage{amssymb}
\usepackage{ulem}
\usepackage{graphicx}
\usepackage{multirow}

\usepackage{cite}

\usepackage{color} 
\def\clrb {\color{blue}}
\def\clrr {\color{black}}

\usepackage[labelfont=bf,labelsep=period,justification=raggedright]{caption}

\makeatletter
\renewcommand{\@biblabel}[1]{\quad#1.}
\makeatother

\date{}

\begin{document}

\begin{flushleft}
{\Large
\textbf{Comparison of modules of wild type and mutant huntingtin and TP53 protein interaction 
networks : implications in biological processes and functions}
}
\\
Mahashweta Basu$^{1}$, 
Nitai P. Bhattacharyya$^{2}$, 
Pradeep K. Mohanty$^{1*}$
\\
\bf{1} Theoretical Condensed Matter Physics Division, \\ Saha Institute of Nuclear Physics, 1/AF Bidhan Nagar, Kolkata 700064, India
\\
\bf{2} Crystallography and Molecular Biology Division, \\ Saha Institute of Nuclear Physics, 1/AF Bidhan Nagar, Kolkata 700064, India
\\
$*$ E-mail: pk.mohanty@saha.ac.in
\end{flushleft}

\section*{Abstract}

Disease-causing mutations usually  change the interacting partners of mutant proteins. 
 In this  article,  we    propose that the biological consequences   of  mutation   
are  directly related  to the alteration of  corresponding  protein protein interaction 
networks (PPIN). Mutation of Huntingtin (HTT)  which causes Huntington's 
disease (HD) and  mutations to TP53  which is associated with 
 different cancers   are studied as  two example cases.   
We  construct the  PPIN of  wild type and mutant proteins separately and identify the structural modules 
of each of  the networks. The functional role of  these  modules are   then assessed by Gene Ontology (GO) 
enrichment analysis for biological processes (BPs).  
We find  that  a large number of significantly enriched ($p<0.0001$) GO terms in  mutant PPIN 
were absent  in the  wild  type PPIN  indicating  the gain of BPs due to mutation.  
 Similarly    some of  the   GO terms enriched in 
wild type PPIN   cease to exist  in  the modules of mutant PPIN,  representing the  loss.
GO terms common in modules of  mutant and wild type networks indicate both loss 
and gain of BPs.
 We further  assign   relevant   biological function(s)  to each  module  by 
classifying the   enriched GO terms  associated  with it.  It turns out that most of   these  
biological functions  in HTT networks are  already known to be altered  in HD  and 
those   of  TP53 networks    are  altered  in cancers. We argue that gain of BPs, and the 
corresponding  biological functions,  are due to new  interacting partners acquired by  
mutant proteins. The methodology we adopt here could be applied to  genetic diseases 
where  mutations alter the ability of the protein to interact with other proteins.


\section*{Introduction}
Cellular functions are carried out by proteins interacting   with other proteins and 
macromolecules like DNA, RNA, etc.    It  is believed \cite{Barabasi} that the  modular organization 
of cellular functions  are related  to the underlying  modular structure  of   the 
protein protein interaction network (PPIN). 
{\clrr Understanding PPIN would elucidate how such interactions execute basic functions in cells and may explain 
the abnormalities arising from mutations in genes.} 
In particular,  mutation at  the binding site of a protein  may lead to loss of 
it's ability to function together with  existing   interacting partner(s). 
On the other hand, mutation  may also create  regions  where   new protein partners can 
bind. Therefore, loss or gain of interaction due to mutation   may  
contribute to causation, progression or modulation of disease. 
It has been reported   recently \cite{Schuster} that 
out of $119$ mutations in $65$ distinct diseases, $95$ mutations result in loss of function (LOF), 
$17$ mutations result in gain of function (GOF) and $4$ mutations changes the preferences for 
interaction. Based on this experimentally validated data, it has been predicted that $1428$ mutations 
might be related to interaction defect.  Using the structural information at 
atomic levels either through crystallography or homology modeling, it has been shown that $21,716$ mutations 
in $624$ genes either alter amino acid sequences or produce truncated proteins.  Among $12,059$ mutations 
that alter amino acid sequences, $7833$ mutations are located in the interface of interaction with 
other proteins.  Such mutations at interfaces of interactions may disrupt or enhance the interactions 
with the partners. This study also emphasizes the role of loss or gain of interactions of mutant 
proteins in human diseases. However, for such analysis, it is necessary to have 
structural information at atomic levels, which may be achieved if 3-dimensional  structures 
of the  proteins   or   their homologs  are known. But, for the most of the protein protein 
interactions such information is not available \cite{Wang}. 
Moreover,   very little is known  about  the role of  such altered interactions in corresponding 
pathological conditions. It remains a challenge to relate genetic mutation data to PPIN and 
to understand molecular cause of disease. In the present communication, we probe whether gain or loss 
of interactions of mutant Huntingtin protein (HTT) that causes Huntington’s disease (HD) can explain  
functional abnormalities observed in  HD. We have also used the same approach to find  how loss or 
gain of interactions of mutant TP53 in cancers may result in  alterations of functions.

\section*{Analysis and Results}

\subsection*{Mutation in HTT protein}
Huntington’s disease (OMIM ID: 143100) is a rare autosomal dominant progressive degenerative 
neurological disease caused by  expansion of normally polymorphic CAG repeats beyond $36$ at 
the exon1 of the gene Huntingtin (HTT) \cite{HD}. Over the years, various cellular 
processes/conditions like excitotoxicity, oxidative stress, mitochondrial dysfunction, endoplasmic 
reticulum stress, axonal transport, ubiquitin proteasome system, autophagy, transcriptional 
deregulation and apoptosis have been  implicated in HD pathology \cite{Imarisio,Ross}.  
 Even though GOF was inferred initially from the autosomal dominant nature of transmittance 
of the disease, the underlying molecular details still remain largely unknown. Inverse 
correlations between age at onset and number of CAG repeat beyond $36$ in HTT gene, increased 
aggregates of mutant HTT (mHTT) and apoptosis, correlation of CAG repeat numbers in HTT gene 
with levels of ATP/ADP and altered expression 
of few genes  \cite{HD,Snell,Seong,Cowan,Jacobsen} suggest  toxic GOF of mutant protein that 
disrupts normal cellular functions and causes neuronal death. Mutant HTT preferentially interacts 
with DNA sequences, alters conformation of DNA facilitating binding of other transcription factors 
to the specific sequences and modulates transcription of genes. This result also indicates a 
dominant GOF of mHTT \cite{Benn}. Wild type HTT (wHTT) is known to be involved in protection of apoptosis 
\cite{Duyao,Nasir,Zeitlin,Rigamonti}, regulation of gene expression \cite{Kegel,ZhangH}, mitosis and 
neurogenesis \cite{Godin}, neuronal development \cite{Tong} and maintenance of body weight 
\cite{VanRaamsdonk}; all these processes are altered in HD \cite{Imarisio,Ross}. These results indicate 
that loss of one of the alleles in HD could contribute to increased apoptosis and altered gene 
expressions observed in HD. LOF of wild type protein may thus  contribute, at least partially, 
to HD pathology \cite{Cattaneo}. There are also several experimental evidences available against 
simple LOF(s) of wild type HTT \cite{Ambrose,Gottfried,Wexler,Myers}.

\subsubsection*{ \it \textbf{Construction of HTT-interacting protein network}}
 We have collected   the HTT interacting proteins from published data  and  find that 
$17$ proteins preferentially interact with wHTT, while $37$ proteins are either 
identified in aggregates of mHTT only or 
interact preferentially with mHTT  (the   references for each of the observations are  provided in  
{\clrb Dataset S1 (sheet 1)} and in {\clrb Supporting Text (Text 1)}).  
These  $17$ and $37$ proteins  are referred to as the primary interactors of wHTT and mHTT respectively.
Next, we assimilate interacting partners of these primary 
interactors from BioGrid (Version 3.1.88, May 2012), a public database that 
contains genetic and protein protein interaction data for humans and other organisms \cite{Stark}. 
In the present study, we have considered both physical and genetic interactions (refer to the section 
`Robustness analysis'  for  details). It turns out that  there are $288$  secondary interactors of wHTT 
(proteins which interact with  the $17$ primary  interactors), whereas there are $1504$ secondary  
proteins which interact  with  $37$ primary interactors of mHTT. The PPIN of wHTT interacting 
proteins is then constructed by considering all these $306$ proteins 
(wHTT + $17$ primary  + $288$ secondary   interactors of wHTT)  as nodes of the network; two 
nodes are connected if corresponding pair of proteins are found to be interacting partners of 
each other in BioGrid.  Altogether there are $1397$ interactions in wHTT network which are 
listed in {\clrb Dataset S1 (sheet 2)}. Similarly the PPIN of mHTT  is constructed with 
$1542$ nodes (mHTT + $37$ primary  + $1504$ secondary   interactors of mHTT)  which has $13142$ 
interactions from BioGrid ({\clrb Dataset S1 (sheet 3)}). We have used  Cytoscape \cite{Shannon} 
for visual presentation of the wHTT and mHTT networks, which are shown in  {\clrb Supporting Text (Fig. S1)}. 
Both the networks are densely interconnected and the nodes are too tangled there to find any apparent 
or obvious modular structures.

 \begin{figure}[h]
\centering
\includegraphics[width=16cm,bb=14 14 2601 900]{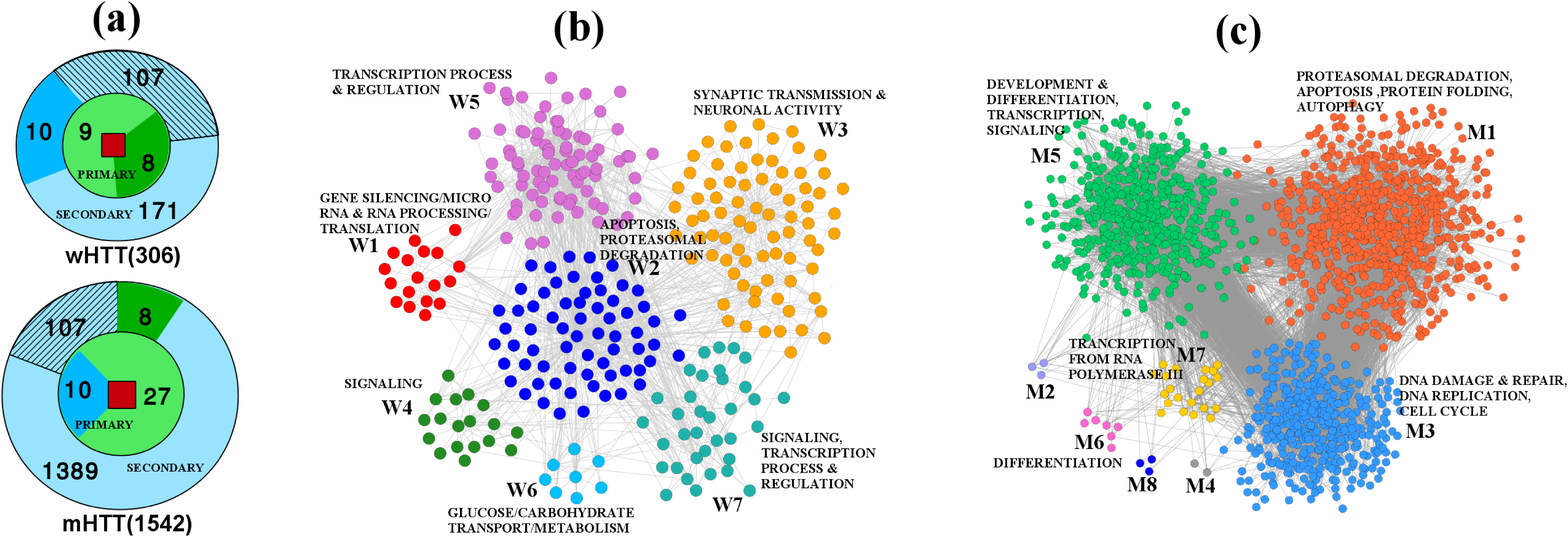}
\caption{ {\bf Construction and modularization of wild type and mutant HTT networks.} \\(a) Proteins involved in the  wHTT  and mHTT
networks: wHTT (mHTT)  protein (red square) has 
$17$ ($37$) primary   and  $288$ ($1504$)  secondary interactors, shown schematically 
as the  inner  and outer  circles. Of the $17$ primary interactors of wHTT, $8$ proteins (deep green) 
become secondary interactors of mHTT.  Among the $288$  secondary interactors of wHTT, $107$ (shaded) 
proteins remain as the secondary interactor  of mHTT whereas  
$10$   proteins (deep blue) becomes the  primary interactors of mHTT.
(b) and (c) Modules of the wHTT and mHTT  networks  from NGM algorithm,
which yields $7$ ($W1, W2,\dots, W7$) and $8$ modules ($M1, M2,\dots, M8$) respectively 
are shown along with the relevant biological functions (obtained GO term  enrichment analysis  from GeneCodis3).  
Significant functions associated with the modules are also shown. Details of the GO terms are shown 
in {\clrb Table S2} and {\clrb Table S3} of the {\clrb Supporting Text}, respectively for wHTT and mHTT.}
 \label{fig:fig3}
\end{figure}

\subsubsection*{\it \textbf{Characteristics of networks}}
 A quantifiable description of these networks can be obtained by using  graph theory, which 
provides several measures for comparison and characterization of complex networks. 
The most elementary characteristic of a node is its degree, $k$, which represents the number 
of other nodes (proteins) it is connected with. The degree distribution, $P(k)$, gives the 
probability that a randomly selected node has exactly $k$ links. We find that both the wild 
and mutant PPINs follow a power law degree distribution, $P(k) \sim k^{-\gamma}$ {\clrb (Fig. S3 } in  { \clrb Supporting Text)}
with exponents $\gamma = 1.99, 1.95$  and   average degrees  $\langle k \rangle = 9.13, 17.05$ 
respectively.  Another important quantity is the  clustering coefficient which characterizes  
how connected are the neighbors of a given node. It is observed that the average  clustering 
coefficient $C=0.361$ for mHTT network is lower compared to $C=0.436$  for wHTT PPIN. This 
indicates that, the former network is less compact and the interacting partners of the proteins 
are poorly connected among themselves. We have also calculated the average shortest path length 
$L$, and the network diameter $D$ (listed in  {\clrb Supporting Text (Table. S2)}), which  describe the  
structural properties of the network. The detailed definitions of $C,~D$ and $L$ along with 
their evaluation procedure is illustrated in  {\clrb Supporting Text (Text 2)}.

\subsubsection*{\it  \textbf{Gain and Loss of interactions due to mutation}}
A closer look at PPINs of wHTT and mHTT reveals that among the $17$ primary interactors of 
wHTT, $8$ proteins still appear in PPIN of mHTT as secondary interactors, $i.e$. they interact 
with some of the primary interactors of mHTT. Again, among $288$ secondary interactors of wHTT, 
$107$ proteins are secondary interactors of mHTT, $10$ proteins interact directly with mHTT and 
the rest $171$ proteins do not take part in PPIN of mHTT (see Fig. 1(a)). Evidently, the mutant 
HTT network has gained several new interactions, $27$ proteins as primary interactors and $1389$ 
proteins as secondary interactors. This result is shown schematically in Fig. 1(a) and the 
detailed list of these proteins is given in  {\clrb Supporting Text (Text 1 and Table S1)}. Since 
mutation of HTT  has changed the PPIN substantially  one expects a significant change in its 
functions.

\subsubsection*{\it \textbf{Modules of wHTT and mHTT networks} }
There are several methods for obtaining  natural modules  of  a network (or partitions of  a graph) \cite{Fortunato}. We adopt Newman-Girvan’s modularization (NGM) algorithm  \cite{Newman}, a commonly used method, 
to detect the modules of wHTT and mHTT networks. This algorithm partitions the network in a way 
that the intra-module connections between nodes  are maximized in comparison to the inter-module 
connections.  To find the modules, Newman and Girvan \cite{Girvan} proposed a score called 
modularity $Q$ for every possible partition of a network; the maximum value of $Q$ corresponds 
to the best partition. The details of the NGM algorithm  for maximization of $Q$ is described 
in  {\clrb Supporting Text (Text 2)}. The NGM algorithm modularizes the PPIN of wHTT into $7$ modules of sizes ($18,~66,~79,~18,~82,~8$ and $35$) (see {\clrb Table S2} in  {\clrb Supporting Text}), with modularity $Q = 0.415$, 
whereas PPIN of mHTT is partitioned into $8$ modules of sizes ($643,~3,~377,~2,~485,~7,~22$ and $3$) 
with $Q = 0.302$. Modules of wHTT and mHTT networks  are denoted by $W$ and $M$ respectively. 
Figures 1(b) and (c) represent the modularized networks; all proteins belonging to a 
given module are shown in same color. Clearly, the mHTT network is visibly more complex 
than that of wHTT, which is consistent with the fact that it has a lower $Q$ value \cite{Qvalue}.

\subsubsection*{\it\textbf{ Similarity between the modules}}
\begin{figure}[h]
 \centering
\includegraphics[width=6.5cm,bb=14 14 720 486]{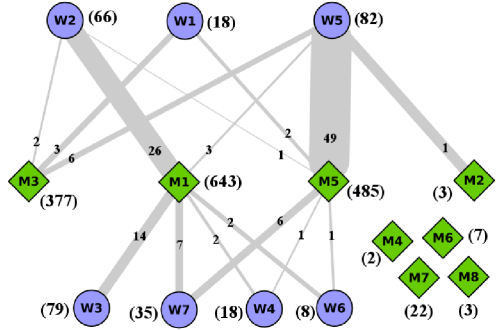}
\caption{ {\bf Similarity between modules of wHTT and mHTT networks.} The figure describes pictorially the closeness between the modules of wHTT and mHTT PPIN; the modules having common protein or common GO terms are joined with edges (numerical value written on the edge as : common proteins).} 
 \label{fig:P_overlapHTT}
\end{figure}

Once the wild type and mutant networks are  modularized, it is important to ask  how similar  
is a  module of wild type network  with that of mutant network, in  terms of their protein 
constituents.  Mutant and wild type HTT networks have $125$ proteins common between them.
After both the networks are modularized, these common proteins are distributed among the pair 
of wHTT- mHTT modules. For example, the module $M5$ ($485$ proteins) has $49$ proteins in 
common with $W5$ ($485$ proteins), whereas it has only one common protein in $W2$ 
(out of $66$ proteins) and two common proteins in $W1$ ($18$ proteins). The  detailed distribution 
of common proteins  among wild and mutant  modules of HTT are shown in  Fig. \ref{fig:P_overlapHTT}.

 To calculate the similarity among  modules, first  we construct a unique set   of proteins from combining 
the  proteins involved in the wild and mutant networks. This set consists of $1723$ proteins 
in case  of HTT. Now every  module of  wHTT and mHTT  are considered as a unique  $1723$ 
dimensional vector as follows. Each protein is identified with a specific position in the vector; 
presence (or absence) of a specific protein in a module, say $w$, is mapped on to a corresponding 
vector $\vec R_w$  by inserting  $1$  (or $0$)   at respective position. A similarity measure 
between a pair of modules $w$ and $m$ is well represented  by the angle  $\theta(w,m)$ between 
the corresponding vectors $\vec R_w$ and $\vec R_m$ , 
$$
\theta(w,m)  =  \cos ^{-1}    \frac{ \vec R_w. \vec R_m} {|\vec R_w| |\vec R_m| } 
$$ 
It is  rather simpler to use  $\cos (\theta(w,m))$ as the similarity  measure  as   $cosine$ function   
is monotonic in the range $(0, \pi).$  It is easy to see that if  the  modules have $N_w$ and $N_m$ proteins  individually  and  $N_{wm}$  protein 
in common, the  similarity measure   is 
\begin{equation}
 \sigma(w,m)  \equiv  \cos (\theta(w,m)) =N_{wm}/\sqrt{ N_w N_m}.
\label{eq:sim}
\end{equation}
Clearly $\sigma(w,m)$ varies in the range $(0, 1)$ with maximum value $1$ corresponding to the 
fact that the modules are identical, $i.e.$ they have same set of proteins.

In Fig. \ref{fig:P_overlapHTT}, we represent the similarity among modules of mHTT and wHTT  as 
a bipartite network with links  having thickness proportional to $\sigma_{wm}$. 
The thickest link between $M5$ and $W5$ indicates that these modules are significantly similar.  For examples, the module $W5$ has $82$ proteins and $M5$ has $485$ proteins; $49$ proteins are common among the proteins in these $2$ modules; thus the  protein similarity index  for $W5$-$M5$ pair is $\sigma(W5,M5)=0.246$. Similarly among $66$ proteins in $W2$ and $643$ proteins in $M1,$ $26$ proteins are common (corresponding $\sigma(W2,M1)= 0.126$).

\subsubsection*{\it \textbf{Enrichment of GO terms for biological process}}
It has been observed that the proteins identified in a particular complex are involved in 
similar functions \cite{Spirin}. From network perspective, these complexes are represented 
by modules and they appear as distinct group of nodes  which are  highly interconnected with each 
other but have only a few  connections with  the nodes outside of the module. It is important to ask, 
if such a structural partition relates to any functional enrichment.  Among   many 
bioinformatics  tools   available for  such  analysis \cite{Huang} we utilize GeneCodis3 \cite{{Tabas}} 
(explained in  {\clrb Supporting Text (Text 3)}) to  obtain the possible Biological processes   enriched    
by   the   proteins  in a  given  module.
{\clrr Given a   query set of proteins GeneCodis3 provides the enriched  biological process, 
molecular functions, and cellular components as defined by the Gene ontology. Biological process 
in Gene ontology is described as a series of events carried out by one or more ordered 
assemblies of molecular functions \cite{GOU}.}
The proteins in each module are used as input to GeneCodis3 \cite{Tabas} and  significantly 
enriched GO terms for BPs obtained using $p$-values 
calculated through Hypergeometric analysis corrected for false discovery rate (FDR). Results of enrichment 
analyses for $7$ modules of wHTT and $8$ modules of mHTT network are shown in 
{\clrb Datasets S2} and {\clrb S3} respectively.

Since many proteins are known to be involved in a particular BP, and a given protein may 
also contribute to multiple BPs, it is likely that proteins in different modules in wHTT 
and mHTT network participate in a specific BP due to either overlap in proteins or BPs. 
To identify the overlaps of BPs between modules in wHTT and mHTT networks, we separately 
identify the common GO terms  between the wHTT and the mHTT modules. It 
is  evident from {\clrb Dataset S4 (sheet 2)} that $390$ unique GO terms are being enriched ($p < 0. 0001$) 
due to proteins in modules of mHTT network, while $129$ GO terms are enriched with proteins in the 
modules of wHTT network ({\clrb Dataset S2 (sheet 1)}). Among the GO terms present in wHTT and mHTT network, 
$65$ are common.  As a result due to mutation, $325$ GO terms are gained by mHTT and $64$ GO 
terms are lost by wHTT. The common $65$ GO terms represents both gain and loss.

{\clrr For convenience, we  clubbed the the GO terms in a  given module   to  broadly assign   
one or more appropriate  biological function(s). For example, GO:0010506 (regulation of 
autophagy), GO:0016559 (peroxisome fission), GO:0031929 (TOR signaling cascade), 
GO:0000045 (autophagic vacuole assembly), GO:0006897 (endocytosis) in module $M1$ are 
bought under  a single biological function ``Autophagy''. Similarly in module $W4$ 
GO:0043507 (positive regulation of JUN kinase activity), GO:0072383 (plus-end-directed 
vesicle transport along microtubule), GO:0046330 (positive regulation of JNK cascade), 
GO:0046328 (regulation of JNK cascade) are clubbed under ``Signaling''.
The assigned  biological functions  for modules  of   wHTT and  mHTT  are shown 
in Fig. 1(b) and (c)  (details   are given in  {\clrb Dataset S4}). }

\subsubsection*{\it \textbf{Gain and loss of biological process  in HTT networks}}
 {\clrr Comparison of enriched  BPs in the modules of wHTT and mHTT reveal that  the 
mHTT network has acquired several new  BPs  which were absent in wHTT,  
indicating  gain  of   biological processes. }
Similarly   enriched BPs  of wHTT   which are  not present in  mHTT  are lost.  
Hence  biological functions carried out by the BPs which are gained or lost in mHTT networks
 may result in functional gain or loss due to mutation in HTT.

\textit{Gain of biological process :} The unique GO terms  enriched 
in the   modules of mHTT networks are listed in {\clrb Dataset S4 (sheet 2)} and in  {\clrb Supporting Text (Table S3)}. 
The GO terms  in   module $M1$ are related to  cell cycle ($4$ GO terms), signaling ($30$), 
transcription processes and regulation ($5$), apoptosis ($11$), DNA damage and repair ($6$), 
Immunological ($7$), protein folding ($7$), autophagy ($5$), translation ($3$), metabolism ($1$), 
development and differentiation ($4$), cell migration and shape ($4$),  proteasomal degradation ($14$), 
Protein complex/membrane assembly/stabilization ($9$) and others ($4$). It is known that many of these processes are involved in HD pathogenesis \cite{Bano}.
 In $M3$, the enriched  GO terms   are   assigned  to  DNA repair ($17$), Transcription processes 
and regulation ($5$), DNA replication ($12$), cell cycle ($12$) and others ($5$). 
 Note that, it has been shown recently  that  DNA repair, replication and cell cycle 
are  involved in HD. In fact, activation of DNA synthesis and cell cycle  increase apoptosis 
in terminally differentiated neuronal cells, instead of increasing cell division \cite{Pelegri,Folch}. 
Besides, recent studies  have explored  the role of DNA repair in neurodegenerative 
disease \cite{Jeppesen} and show that interaction of mHTT with Ku70/ XRCC6 impairs 
repair activity \cite{Enokido}. 
A large number of GO terms related to  development and differentiation ($57$ GO term), transcription 
process and regulation ($31$), cell cycle ($5$), DNA damage and repair ($4$), 
Carbohydrate/Glucose transport/metabolism ($4$), Cell growth ($7$), signaling ($31$) and others 
($6$) are enriched in module $M5$. The role of development and differentiation in HD is not clear. {\clrr However recent 
studies  in HD \cite{Humbert,Tong}   indicate that neurogenesis is possibly altered  and 
differentiation/development could be defective.} Deregulation 
of transcription is considered to be one of the most important abnormalities in 
HD \cite{Seredenina}. GO terms related to differentiation are also enriched with proteins 
in module $M6$, although the terms are distinct from that in module $M5$. All $4$ GO terms 
enriched in $M7$ are related to transcription by RNA polymerase III. 
It is known that both tRNA and some miRNAs \cite{Borchert} are synthesized by RNA polymerase III, 
however their role in HD is unknown. Thus it is evident that  the   the protein interactions  
gained  in  mHTT  network result in enrichment of  the biological processes  in  its  modules.

 \textit{Loss of biological process :} The unique GO terms  enriched in the 
modules of wHTT which are absent in the modules of mHTT network represent the loss of 
functions due to mutation in HTT protein. The $5$ GO terms in $W1$ include gene silencing, 
micro RNA processing and translational regulation. The GO terms  relating to  proteasomal 
degradation ($1$ GO term), cell cycle ($1$), apoptosis ($1$) and circadian rhythm ($1$) 
are present in $W2$. Similarly, signaling ($17$ GO terms), synaptic transmission, neuronal 
activities ($12$) transport (ion/sugar) ($5$) and others ($4$) are associated with module $W3;$ 
glucose/carbohydrate transport and metabolism ($5$), cell cycle ($3$) and protein/transmembrane 
transport ($4$) with $W6.$ In $W7$ only one GO term describing transcription processes and 
regulation is enriched. The GO terms and the associated BPs that are lost due to mutation are provided 
in {\clrb Dataset S4 (sheet 1)} and in  {\clrb Supporting Text (Table S3)} respectively.

We  have clubbed the relevant GO terms to represent signaling, 
transcription process and regulation, apoptosis, cell cycle etc. (refer to {\clrb Table S3} of 
 {\clrb Supporting Text}). 
For example, GO terms (GO:0000088) and (GO:0000236, GO:0000087, GO:0007091) which are enriched in $W2$ 
and $W6$ respectively relates to cell cycle. Similarly the $21$ GO terms which are enriched 
in $M1(4),$ $M3(12),$ and $M5(5)$ ({\clrb Dataset S4 (sheet 2)}) are also associated to cell cycle. 
Although cell cycle  is enriched in both wHTT and mHTT modules, no 
 GO terms are common among them. Thus, the loss of interaction with wHTT may result in loss of 
above $4$ GO terms in wild type network resulting in LOF, whereas the gain of interaction 
with mHTT may be associated with gain of these $21$ GO terms relating to GOF of cell cycle.

It is interesting to note (from {\clrb Table S3} of  {\clrb Supporting Text}) that the GO terms related to DNA 
replication, protein folding, autophagy, cell growth are only observed in the modules 
of mHTT networks. So these processes are gained due to new interaction with mHTT. 
Similarly, GO terms related to gene silencing/ microRNA processing/ translation, transport 
(ion/protein/sugar etc) are observed in wHTT network only. Therefore, loss of interaction with 
wHTT may result in the loss of these BPs in HD.

\textit{ Both loss and gain of biological process :} {\clrr Modules in wHTT and mHTT networks   have 
several proteins or GO terms   common among them, which  indicate  loss as well 
as gain of  functions  and  support the notion that both loss   and  gain may occur  
due to mutation in HTT \cite{Cattaneo}.} For example, modules ($W1, W5, W7$) and 
($M3, M5, M6$) have $17$ enriched GO terms related to transcription processes and regulation. 
Similarly, modules ($W2, W5$) and ($M1, M3, M5$) share $12$ enriched common GO terms related 
to apoptosis and $4$ common GO terms relating to cell cycle.  Thus, the general function of 
transcription and apoptosis could arise from loss as well as gain of interactions of mHTT 
protein. {\clrr The details of the functions associated with the $65$ GO terms (common between wHTT and mHTT)
are presented  in {\clrb Dataset S4 (sheet 3)} and in  {\clrb Supporting Text (Table S3)}, they correspond to the gain and loss of functions in the HD.}

 From the above analysis we observe that most of the functions that are enriched in the 
modules of wHTT and mHTT networks are altered   in the pathogenesis of  HD.
The post transcriptional regulation of genes, associated 
with module $W1$ of wHTT network, can be related to negative regulation of gene expression by 
the non-coding RNAs like micro RNAs, which are well documented \cite{Sinha}. Role of apoptosis 
\cite{Imarisio,Bano}, synaptic transmission \cite{Milnerwood}, JNK pathway \cite{Perrin}, 
transcription deregulation \cite{Seredenina}, glucose transport \cite{Gamberino,Ciarmiello}, 
estrogen \cite{Bode} and various types of epigenetic changes including histone modifications 
in different neurological diseases \cite{Urdinguio} in HD pathogenesis have also been reported.

In summary, many new BPs (GO terms) appear  in the    mHTT network and 
some of   the  BPs   present in wHTT network  are lost; a  few are   found to be common  
between  modules of wHTT and mHTT. As a result some biological functions involving the enriched  
GO terms are gained  by  mHTT  and a  few  are lost   from the  modules of wHTT. 
This provides molecular mechanism of  the gain and/or loss of functions 
observed in HD pathogenesis.

\subsection*{Mutation in TP53 protein}

TP53 protein, initially identified as an oncogene,  is now established  as a tumor suppressor 
gene   which  participates in diverse cellular functions like transcription regulation, 
DNA repair, apoptosis, and genome stability, and many others. Mutation   to TP53  is 
identified in more than $50\%$ of the tumors. It is evident from COSMIC database \cite{Forbes}  that R175H, 
R273H and R248W mutations  of TP53 are  the  most prevalent ones.
 Since TP53 is  a tumor suppressor gene,  it  is expected  that its  mutations might result in the LOF 
of the wild type protein. Some  mutations  of TP53  are also known to attain 
new function(s) \cite{Freed-Pastor,Oren}. For example, exogenous expression of mutant TP53 
(R273H and others) in mouse cells devoid of endogenous TP53 results in several cellular 
phenotypes of cancers \cite{Dittmer,Adorno,Muller}. To understand the underlying molecular 
mechanism of GOF of mutant TP53, it was recently shown  \cite{CoffillCR} that nardilysin (NRD1) protein, 
which does not interact with wild type TP53 but interacts  only with mutant TP53 (R273H),  
may contribute to the metastatic properties  of this mutant protein.

\subsubsection*{\it \textbf{PPIN of wTP53 and R273H mutant TP53 (mTP53)}}

In a recent study  \cite{CoffillCR},  it  has been  shown  that $17$ proteins 
preferentially interact with the wild type TP53 (wTP53) and $30$ other proteins interact exclusively 
with  mutant TP53 (mTP53).   {\clrr  To construct   the     protein interaction networks 
we   take  these primary  interacting proteins  of  wTP53 and mTP53  and   consider their  interacting 
partners  existing in BioGrid database \cite{Stark}. } The detailed protein interaction data are 
given in the {\clrb Dataset S5}. The PPIN is constructed separately for wTP53 and mTP53,  as described 
 for HTT. It turns out that wTP53 has $601$ secondary interactors whereas mTP53 has only $547.$  
Thus the PPIN of wTP53 and mTP53 are constructed taking
$619$ proteins $({\rm wTP53} +17\;{\rm primary} + 601\;{\rm secondary})$ and 
$578$ proteins $( {\rm mTP53} + 30\;{\rm primary} + 547\;{\rm secondary})$  respectively. 
{\clrr Both the networks (shown in  {\clrb Supporting Text (Fig. S2)}) are  found to be  densely packed   with similar
structural properties.}  Their degree distributions are scale free ($P(k) \sim k^{-\gamma}$) with the 
exponents $ \gamma = 2.04$ (wTP53) and $1.89$ (mTP53) ({\clrb Fig. S3} in  {\clrb Supporting Text}) and average 
degree $\langle k \rangle = 12.07, 12.29.$ The other network properties, like the average 
clustering coefficient $C=0.452,0.406$, the diameter of the networks  $D=4,4$ are also 
comparable (listed in  {\clrb Supporting Text (Table S2)}).

\begin{figure}[!ht]
\centering 
\includegraphics[width=16cm,bb=14 14 2505 945]{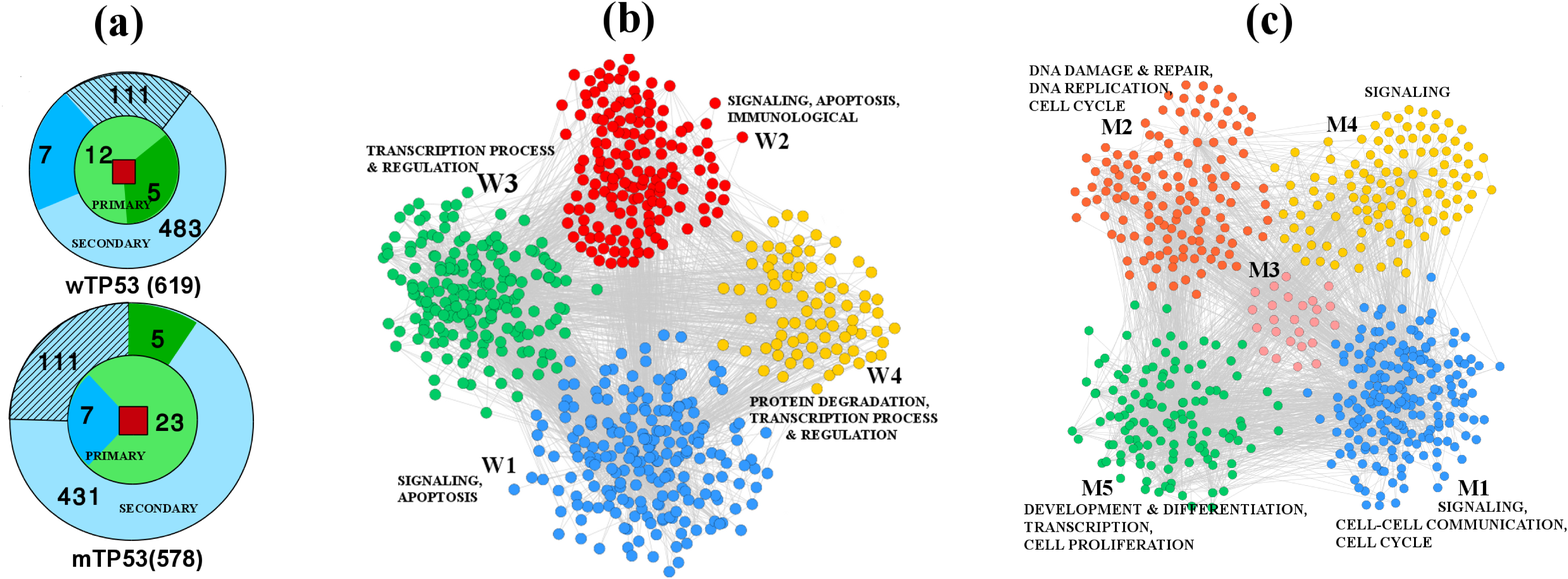}
\caption{{\bf Construction and modularization of wild type and mutant TP53 networks.} \\(a) Proteins in   wTP53  and mTP53 networks :
wTP53 (mTP53)  protein (red square) has $17$ ($30$) primary   and  $601$ ($547$) secondary interactors,  
represented  by the  inner  and outer  circles.  Only $5$ ($7$) primary (secondary)   proteins of wTP53 
interact  with mTP53 as secondary (primary) interactors.  Again $111$ secondary proteins of wTP53  
remain as   secondary interactors of  mTP53.  (b) and (c)  shows the modules of  wTP53 and mTP53 network 
along with few  plausible candidate BPs. Details of the GO terms are shown for wTP53 and mTP53 respectively in 
{\clrb Table S6} and {\clrb Table S7} of the {\clrb Supporting Text}.}
\label{fig:fig7}
\end{figure}

The change in interactions and the interacting partners due to mutation of TP53 is shown schematically 
in Fig. \ref{fig:fig7}(a). Of $17$  primary  interactors of  wTP53, only $5$ proteins remain 
involved in mutant network as secondary interactors of mTP53 and the  remaining $12$  do not 
interact with mTP53. Among the $601$ secondary interactors of wTP53,  $111$ proteins remain as a 
secondary interactor of mTP53  and  $7$ of them  interact  directly, $i.e.$ $7$ secondary interactors 
of wTP53  become  primary interactors of mTP53.  Lists of these proteins are given in  {\clrb Supporting Text (Text 1
 and Table S1)}.

\subsubsection*{\it \textbf{Modules of wTP53 and mTP53 networks}}
\begin{figure}[h]
 \centering
 \includegraphics[width=5cm,bb=14 14 632 584]{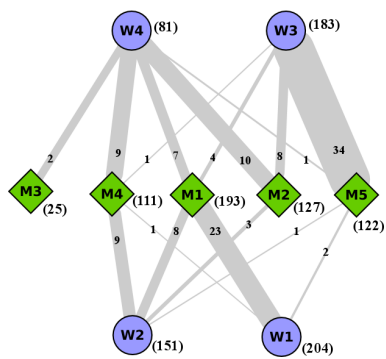}
\caption{{\bf Similarity between modules of wTP53 and mTP53 networks.} The bipartite network constructed with the  modules of wTP53 and mTP53; the common proteins present between a pair of wild and mutant module is written on respective link. The number of proteins that constitute the modules are written beside it.}
 \label{fig:P_overlapTP53}
\end{figure}

In order to identify the modules of the  wTP53 and mTP53 networks, we use NGM algorithm \cite{Newman}. 
It turns out that PPIN of wTP53 is modularized into $4$ modules of size $204, ~151, ~183$ and $81,$ 
whereas mTP53 network gives $5$ modules of size $193, ~127, ~25, ~111$ and $122.$ The corresponding 
modularity values are $Q =0.331$ and $0.338.$ Figures \ref{fig:fig7}(b) and  (c) show the modules 
of wTP53 and mTP53  with different colours.  Each module of wTP53 or mTP53 has unique set of 
protein. However, there is a large overlap of secondary interactors (proteins which do not interact 
directly with TP53) in the wTP53 and mTP53 networks, which is distributed among different modules 
(in total $123$).  We observe that among $123$ common proteins, $34$ belong to module $W3-M5$, 
whereas module pairs $M1-W1$ (and $W4-M2$) have $23$ (and $10$) common proteins. One can define a 
similarity measure $\sigma_{mn}$ using Eq. (\ref{eq:sim}) for every pair of wTP53-mTP53 modules. 
Taking the similarity indices $\sigma_{wm}$ as weights (or thickness) of the link we have constructed 
a bipartite network which is shown in Fig. \ref{fig:P_overlapTP53}; the number of  proteins  is written 
beside each of the modules and the number of common proteins is specified along the links.

Enrichment of biological processes for the proteins present in every module of wTP53 and mTP53 
PPIN using GeneCodis3 are presented in {\clrb Dataset S6} and {\clrb S7} respectively, where only 
the GO terms with $p<0.0001$ are considered. The number of enriched GO terms in modules of 
wTP53 PPIN are $W1(30),\;W2(17),\;W3(63),\;W4(36)$ and  those   for  mTP53   are $M1(52),\;M2(71),\;M4(1),\;M5(67).$ 
Note that module $M3$ has no GO terms enriched with $p<0.0001.$ 


\subsubsection*{\it \textbf{Loss and Gain of  biological processes in TP53 networks}}
Enrichment analysis  of  proteins in modules of wTP53 and mTP53  using GeneCodis3 reveals
that respectively $127$ and $172$ GO terms (or biological processes) are enriched 
significantly $(p < 0.0001).$   Among $127$   GO terms   of  wTP53   $57$  GO terms 
do not appear in the mTP53 representing loss of the corresponding biological processes. 
Again the   mTP53    network  has   $102$ new GO terms  (which were absent in wTP53).     
Besides,  $70$ enriched GO terms  are found to be common in  modules of  mTP53 and 
wTP53 networks. {\clrr We further associate each of the enriched  GO terms with a relevent function.
Loss and gain of these broadly  classified functions are discussed below.}

 \textit{Gain of biological processes :}  The  biological processes  related to $102$ 
 new GO terms  of  mTP53  are  gained due to mutation.      
The functions  enriched
in module $M1$ of mTP53 network are  cell-cell communication (no of GO terms $5$), signaling ($13$), 
protein complex/membrane assembly/stabilization ($4$), proteasomal degradation ($2$), cell cycle ($3$), DNA damage and repair ($1$) 
and others ($2$). GO terms related to DNA 
replication ($11$), DNA damage and repair ($14$), cell cycle ($4$), immunological 
functions ($3$), proteasomal degradation ($3$) and signaling ($1$) are enriched in $M2$. 
Similarly GO terms related to differentiation and development ($11$), signaling ($7$), transcription ($6$) , 
cell proliferation ($4$), apoptosis ($1$), cell cycle ($1$) and DNA damage ($2$) and others ($4$) 
are enriched with proteins in module $M5$. The extensive list of the GOF is given 
in {\clrb Dataset S8 (sheet 2)} and in  {\clrb Supporting Text (Table S4)}. Thus new functions carried out by these biological
 processes are due to gain of interaction.

 \textit{Loss of biological processes :} On the other hand some   of the enriched  GO terms of   wTP53 
 are   absent in  the  mutant network. Corresponding   biological processes  are 
lost due to mutation in TP53. Altogether  $57$ unique GO terms are enriched with proteins in 
modules of wTP53 networks  which   are classified   into  broad class of functions 
(see {\clrb Dataset S8 (sheet 1)} and  {\clrb Supporting Text (Table S4)}). The  resulting  
loss of biological functions  in   various modules are,   $W1:$ 
signaling ($8$), proteasomal degradation ($1$), translation ($1$), cell migration and 
movement ($2$) and others ($1$);  $W2:$   signaling ($4$), apoptosis ($2$) and
immunological ($3$);  $W3:$ cell cycle ($3$), signaling ($1$), transcription process and regulation ($13$), DNA replication ($3$), DNA damage and repair ($1$); $W4:$ transcription process and regulation ($2$), proteasomal degradation ($2$), translation ($5$) and others ($4$); $W7:$ transcription process and regulation ($1$). 

\textit{Both loss and gain of biological processes :}  The $70$ GO terms common  between wTP53  and mTP53 networks are related to the functions, cell cycle ($14$ GO terms), transcription ($15$), DNA damage and repair ($10$), cell growth ($2$) and apoptosis ($4$), signaling ($9$), DNA replication ($3$), proteasomal degradation ($3$), immunological ($2$), development and differentiation ($1$), metabolism ($2$) and others ($5$).  {\clrr Thus these  functions  are possibly enriched due to both gain and loss of interactions  (details are shown in {\clrb Dataset S8 (sheet 3)} and in  {\clrb Supporting Text (Table S4})).}

\subsubsection*{\it \textbf{Analysis of proteins in different modules using tool GeneDecks}}

 Recently metastasis has been shown as the GOF as R273H cells attain metastatic property in cell 
model \cite{CoffillCR}. Since metastasis is not described as a ``biological process'' in Gene 
Ontology term, we have used another tool, GeneDecks \cite{Safran}, which provides a similarity 
metric by highlighting shared descriptors between genes, based on annotation within the GeneCards 
compendium of human genes (see {\clrb Text 4} in  {\clrb Supporting Text}  for details). Taking the proteins of the modules 
of wTP53 and mTP53 separately as a query field, we look for ``metastasis'' in the attribute ``disorder'' 
among many other descriptors which are enriched for different types of cancers ({\clrb Dataset S9}). 
It is observed that the descriptor ``metastasis'' is enriched with the protein modules $W1,~W2,~W3$ 
of wTP53 network and all the modules ($M1,~M2,~M3,~M4,~M5$) of mTP53 network.  Thus, the loss of 
interactions of proteins in the modules $W1,~W2,~W3$ of wTP53 due to mutation may result in the 
LOFs related to metastasis. Similarly, the gain of interactions of proteins in all the modules 
of mTP53  may result in the GOFs related to metastasis. 


 That LOF of wTP53 and GOF of mTP53 may contribute to invasion and metastasis, 
is reviewed recently \cite{Muller2011}.  TP53 mutations at the DNA binding domain are 
common and such mutations suppress expression of target genes. It is supported by 
several experiments \cite{Muller2011}  that suppression of transcriptional 
program for genes involved in epithelial-mesenchymal transition (EMT) may contribute to 
induction of EMT  resulting in metastasis.
Further, it is ascertained  that loss of functions in wTP53 lead to
increased cell motility in various cell types, and increased 
expression of fibronectin, collagens and extracellular matrix (ECM) proteins. Enhanced 
expression of these proteins potentially increase the interaction between cells and ECM. 
LOF in  wTP53   also  activate Rho GTPases and modulates cell migration \cite{Muller2011}. 

Role of mTP53 in metastasis  has been  established  in many  other studies. 
 Mutant TP53 (R175H) is involved in TGF mediated invasion and metastasis in breast cancer cells through TP63 and SMAD3 \cite{Adorno}. Note that, in our analysis,   SMAD  is present in  module $M5$  of mTP53 network.  
 It is known that  mutant TP53 (R175H and R273H) increases endocytic recycling of adhesion molecule 
integrin and EGFR promoting and metastasis \cite{Muller,Selivanova}.  Mutation in TP53  also 
activate EGFR/PI3K/AKT pathways   and thereby increases invasion \cite{Dong}. Various 
other mechanisms  of increased metastasis by the mutant TP53 have also been studied \cite{Muller2011}.
Thus  the gain of   biological processes obtained from the analysis   of   mTP53  protein networks 
provides an explanation of GOFs observed  in cancers.

\subsection*{Robustness analysis}
In general,  the modularization   methods 
partition  the network  into  communities  of  proteins  which are densely connected. Thus in a large 
network   it is quite expected that deletion  of   small fraction of links, whether   selected  
methodically or   randomly,  does not  alter the  overall   structure  significantly. 
In fact, the degree distributions of   all   four networks studied here (namely  PPIN of wHTT, mHTT, 
wTP53 and mTP53) are scale free  (see  {\clrb Fig. S3} of  {\clrb Supporting Text}), 
and it is known that such scale free networks are 
robust against random removal of nodes or links, but they could be fragile 
against targeted attack \cite{AlbertNature}.

\begin{table}[h]
  \caption{Change in the total number of proteins and the  interactions after excluding 
(a) genetic interactions and then (b) excluding interactions which are validated by only one Y2H experiment.}
\label{table:robust1}
\begin{center}
 \footnotesize  { \begin{tabular}{|c || p{1.8cm} p{2.2cm} p{2.7cm} | p{1.8cm} p{2.7cm}|}
\hline
 & 
{\bf Total no. of  interactions} & {\bf (a)Excluding genetic(\%)}& {\bf (b)Excluding genetic \& Y2H(\%)} & {\bf Total no. of proteins} & {\bf Excluding genetic \& Y2H(\%)} \\
\hline
{\bf PPIN(human) } & $59027$ & $58927(99.84\%)$ & $47244(80.04\%)$ & $12515$ & $11630(9.29\%)$ \\
       {\bf  wHTT} & $1380$  & $1375(99.64\%)$  & $1231(89.20\%)$  & $306$   & $292(9.54\%)$   \\
       {\bf  mHTT} & $13105$ & $13058(99.64\%)$ & $11478(87.58\%)$ & $1542$  & $1486(9.64\%)$  \\
      {\bf  wTP53} & $3718$  & $3716(99.95\%)$  & $3205(86.20\%)$  & $619$   & $590(9.53\%)$   \\
     {\bf   mTP53} & $3521$  & $3515(99.83\%)$  & $3136(89.07\%)$  & $578$   & $551(9.53\%)$   \\
\hline
 \end{tabular}
}
\end{center}
\end{table}

Again, since  several databases of protein interactions largely 
overlap \cite{databases} in their contents,   it is  natural   to expect  that  
the broadly classified biological functions obtained here  for HTT and TP53 
networks  would not differ substantially. In this study 
we used Biogrid \cite{Stark} for creating the  differential PPIN  of the  wild type and 
mutant HTT and TP53 proteins by connecting  every pair of proteins which are
listed  in BioGrid as interacting partner of each other. This includes experimentally validated   
 genetic and physical interactions. 
To check the robustness of our analysis, first let us remove all genetic 
interactions  listed in BioGrid. This reduces  the  total  number 
of protein interactions   of   BioGrid 
to $99.84\%$, whereas the interactions of wHTT, mHTT, wTP53 and mTP53 are reduced 
to $99.64\%$, $99.64\%$, $99.95\%$ and $99.83\%$ respectively (see Table \ref{table:robust1}).  
Among the other experiments  considered in  BioGrid,  Yeast $2$ Hybrid (Y2H) 
assay results in larger  false positives \cite{FDR}. Thus we further remove all the 
interactions  which are  identified only once  by Y2H. This stringent  criterion consequently 
reduces both the  number of interactions and  the  number of  proteins by $ \sim 10\%.$  
The total number of   interactions  of BioGrid  is, however, reduced by $20\%$.  
Since the  wild type and mutant networks are  altered  only a little compared to 
the expected   value $20\%$, one expects   that   deletion of a small fraction of 
interactions   will not change   the network properties significantly.

\begin{table}
\caption{
Comparison of number of proteins and GO terms in the modules of mHTT with respective of `most similar module'
 of the network (a) after excluding genetic and Y2H experiments and (b) after deletion of $10\%$ links.}
\label{table:robust2}
\begin{center}
\footnotesize {
\begin{tabular}{|c||r r||r|c||r|c|}
\hline
\multirow{2}{*}{} & \multicolumn{2}{|c||}{ {\bf mHTT Module}} & \multicolumn{2}{|p{2.5cm}||}{{\bf (a) Excluding genetic \& Y2H}} &  \multicolumn{2}{p{3.3cm}|}{ {\bf (b) Random deletion of  $10$\% links}} \\
\cline{4-7}
& &  &  MSM & Common ($\%$) & MSM & Common ($\%$) \\
\hline
\multirow{2}{*}{{\bf M1}} & No. of Proteins :& $643$ & $656$ & $542~(84.29\%)$ & $612$ & $521~(81.03\%)$ \\
& No. of GO  terms :& $161$ & $161$ & $147~(91.30\%)$ & $163$ & $147~(91.30\%)$ \\
\hline
\multirow{2}{*}{{\bf M3}} & No. of Proteins :& $377$ & $287$ & $211~(55.97\%)$ & $309$ & $209~(55.44\%)$ \\
& No. of GO terms :& $78$ & $95$ & $64~(82.05\%)$ & $80$ & $59~(75.64\%)$ \\
\hline
\multirow{2}{*}{{\bf M5}} & No. of Proteins :& $485$ & $442$ & $423~(87.22\%)$ & $424$ & $397~(81.86\%)$ \\
& No. of GO terms :& $198$ & $186$ & $173~(87.37\%)$ & $174$ & $162~(81.82\%)$ \\
\hline
\multirow{2}{*}{{\bf M6}} &No. of  Proteins :& $7$ & $5$ & $5~(71.43\%)$ & $5$ & $5~(71.43\%)$ \\
& No. of GO terms :& $12$ & $16$ & $11~(91.67\%)$ & $16$ & $11~(91.67\%)$\\
\hline
\multirow{2}{*}{{\bf M7}} & No. of Proteins :& $22$ & $7$ & $7~(31.82\%)$ & $612$ & $7~(31.82\%)$ \\
& No. of GO terms :& $5$ & $8$ & $5~(100.0\%)$ & $7$ & $4~(80.00\%)$\\
\hline
\end{tabular}
}
\end{center}
\end{table}

To demonstrate this explicitly,  we reconstruct the PPIN of mHTT keeping only the 
reduced set of interactions and then identify   the  protein modules  using  Newman Girvan 
algorithm. The   enriched  GO terms ($p<0.0001$)  from GeneCodis3 shows   
that  every  module  of mHTT  ($M1,~M3,~M5,~M6$ and $M7$) has  significant protein 
overlap with only `one  distinct module'  of the reduced network, which is  
referred  to as the `most similar module' (MSM)  henceforth.  The  number of   
overlapping  proteins and GO terms   between the  modules of mHTT and  their 
corresponding  MSM in the reduced network are listed  in Table \ref{table:robust2}.
Evidently, in all cases, about $90\%$   of the  GO terms are retained.  
Thus, the  loss, gain and  loss/gain of biological processes  obtained  
from BioGrid are quite robust. 
 
For  completeness, we  also removed randomly $10\%$ links of mHTT network   
and   repeat the  above analysis which is summarized in Table \ref{table:robust2}.   
Again,  we find that about $90\%$ of the  GO terms   enriched in this network 
are identical to those   obtained for mHTT. Thus, in general,  the enriched biological 
processes  obtained through this analysis are quite robust.

\section*{Discussion and Conclusion}
\begin{figure}[h]
 \centering
 \includegraphics[width=12cm,bb=0 0 794 595]{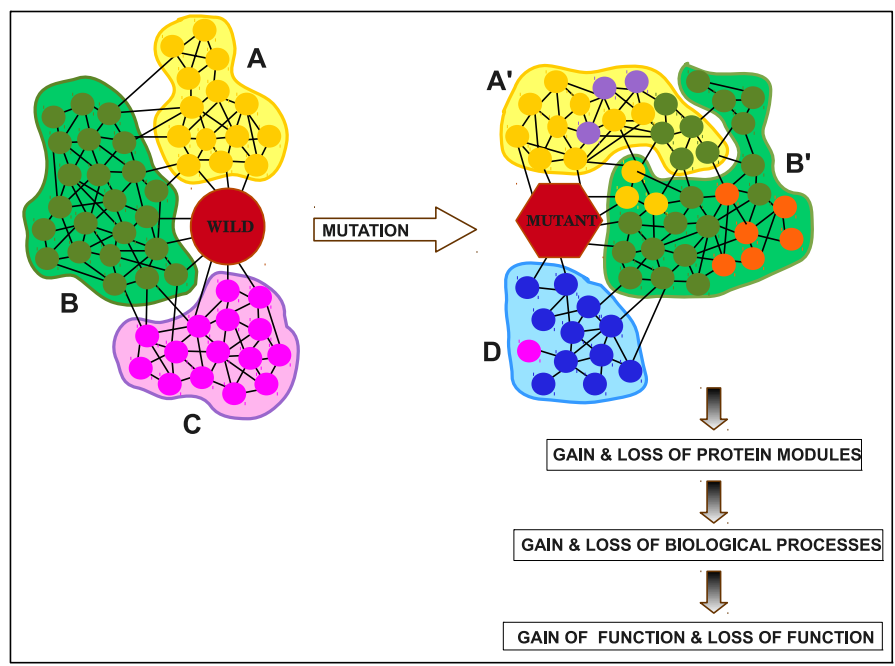}
 \caption{ {\bf Loss and gain of functions from differential network studies.}
The general philosophy of the current work is described here for a schematic  
protein interaction network, where the wild type and the mutant protein have $52$ and $54$ interactors respectively.
There are three modules in 
wild type  network ($A$, $B$, and $C$); all proteins  in a given module are marked   with same colour.
After mutation   the  protein looses  some interactors (marked as pink) and gain some 
new ones (marked as orange, blue and violet). The  PPIN of mutant protein  has three modules $A'$, 
$B'$ and $D$. Module $A'$, which primarily contains proteins of module $A$, has some proteins 
from  module $B$ (green) and some new proteins (violet).  
Most of the proteins in module $D$ are  new interactors  and  thus this module is  
gained by the mutation.  Similarly   proteins  of module $C$  have lost  their interactions.   
Correspondingly, the BPs  which are enriched in  module $D$  are  
gained   and    those   enriched in  module $C$  are lost. This 
loss  or gain of   BPs  lead to  loss or gain of   functions  
in the  pathogenesis of  the  mutation  causing disease.}
 \label{fig:schema}
\end{figure}

Mutation in  protein may change its preference for binding with other proteins and alter 
the corresponding PPIN substantially. We use a graph theory based modularization approach 
to identify the modules of PPINs, and  provide a comparative study of these differential 
networks  using  two examples; one for HD and another for cancers. The general philosophy 
of this analysis   is depicted  schematically in Fig. \ref{fig:schema}. 
In this figure, the wild type protein  interacts with   many  other proteins 
forming a complex interaction network. Broadly,  the schematic wild type network 
has three subgraphs or modules ($A,$ $B$ and $C$); proteins  in each module are marked there with  
identical colours.  The mutant protein loses  some   proteins as  interacting 
partners  (marked as pink) and gains some new ones (marked as orange, blue and violet). 
The network of the mutated protein  has a  revised  modular structure $A'$, $B'$ and $D$. 
Module $A'$  and $B'$ are re-structured and they   have  some proteins from  other modules  
and some new proteins.  Module $D$  is gained by the mutation as  most of  proteins in 
this module  were not present in the wild type network,   and  module $C$ is lost.
Correspondingly, the biological processes  (GO terms) which are enriched in  module $D$  are  
gained   and    those   enriched in  module $C$  are lost.   We argue that  this 
loss  or gain of   BPs  lead to  loss or gain of   functions  
in the  pathogenesis of  the  mutation  induced disease. 
  
In this article  we  explained the   general  idea  of `obtaining  the  loss  
and gain of functions  from  the loss and  gain    of  BPs   enriched in protein  modules'
using two examples; one for HD and another for cancers.
Our analysis predict  a set of broadly classified  biological processes 
(from  the the  GO terms enriched  
in the modules of HTT and  TP53 networks) which    could be  involved in  the  
pathogenesis  of HD  and cancers respectively. In HD, the broadly classified BPs, like 
 post transcriptional regulation of genes, apoptosis, synaptic transmission,  JNK pathway, 
transcription deregulation,  glucose transport, histone modifications etc are 
enriched with the proteins in modules of wHTT and mHTT networks. These BPs 
are already known to be altered in HD pathogenesis.  Similarly, the 
gain and loss of BPs mTP53   results  in the metastatic properties, 
which  have been observed recently.  

Although, we demonstrated  the plausible  loss and gain of biological processes in 
two  examples where mutation alters protein interaction networks of wild type protein, 
the methodology discussed   here  can be  adopted and applied to study 
differential  PPIN in  general.  In particular,  knowing  the  changes in  the protein 
interaction network, either due to mutations that modify   the structure of the 
protein at the binding surface or due to the change in interaction environments,   one can 
predict  what alteration  might occur  in   the   biological processes and functions.  
Such analysis may help understanding the  loss  or gain  of biological processes/functions 
in   genetic diseases caused by mutations. This may in future lead to better 
design of disease intervention through targeting the biological processes/functions 
of specific modules.

\section*{Supporting Information} 
\begin{itemize}

\item[]  \textbf{Supporting Text :} 
Text $1$: Differential interaction due to mutation in HTT and TP53.\\
Text $2$: Analysis of network structure.
Text $3$: Enriched biological processes in modules.\\
Text $4$: Enrichment of metastasis  from GeneDeck. (PDF) 

 \item[] \textbf{Dataset  S1 :} Differential interaction of the wHTT  and mHTT protein. (XLS) 

\item[]  \textbf{Dataset  S2 :} The proteins  belonging  to different modules of wHTT network 
and    their GO term enrichment analysis. (XLS) 

\item[]  \textbf{Dataset  S3 :}  The proteins  belonging  to different modules of mHTT network 
and  their GO term enrichment analysis. (XLS) 

\item[]  \textbf{Dataset  S4 :} The list of LOF,GOF and GOF/LOF for wHTT and mHTT networks. (XLS) 

\item[]  \textbf{Dataset  S5 :} Differential interaction of the wTP53  and mTP53
 protein. (XLS) 

\item[]  \textbf{Dataset  S6 :} The proteins  belonging  to different modules of wTP53 network  
and  their GO term enrichment analysis. (XLS) 

\item[]  \textbf{Dataset S7 :} The proteins  belonging  to different modules of mTP53 network 
and   their GO term enrichment analysis. (XLS) 

\item[]  \textbf{Dataset  S8 :} The list of LOF,GOF and GOF/LOF for wTP53 and mTP53 networks. 
(XLS) 

\item[]  \textbf{Dataset  S9 :} The GeneDeck analysis  of the  proteins  in  the modules of wTP53 and mTP53 networks and enrichment of  metastatsis. (XLS) 

\end{itemize}

\section*{Acknowledgments} 
The authors  acknowledge Saikat Mukhopadhyay for his technical help 
and Urna Basu for careful reading of the manuscript.


\end{document}


\begin{flushleft}
{\Large
\textbf{Supporting Information}
}
\\
Mahashweta Basu$^{1}$, 
Nitai P. Bhattacharyya$^{2}$, 
P. K. Mohanty$^{1}$
\\
\bf{1} Theoretical Condensed Matter Physics Division, \\ Saha Institute of Nuclear Physics, 1/AF Bidhan Nagar, Kolkata 700064, India
\\
\bf{2} Crystallography and Molecular Biology Division, \\ Saha Institute of Nuclear Physics, 1/AF Bidhan Nagar, Kolkata 700064, India
\\
\end{flushleft}

\section*{Text 1 : Differential interaction due to mutation of HTT and TP53}
 \subsection*{ Huntingtin protein} 
The   mutated huntingtin protein mHTT preferentially interact with a set  of different proteins compared to 
the wild type HTT (wHTT). The data for  this differential interactions  are curated from   literature 
as follows. Primarily we have collected $328$ HTT-interacting proteins identified by yeast $2$ hybrid (Y2H) 
assays \cite{Faber,Goehler,Kaltenbach}, affinity pull down followed by mass spectrometry assay \cite{Kaltenbach} and 
proteins in the aggregates of mutant HTT \cite{Mitsui}.  Since Y2H assay results in false positives 
\cite{VonMering}, we consider only $154$ HTT interacting 
proteins that are validated by (i) second method like co-immunoprecipitation or (ii) the
interacting proteins are tested for their implications in HD pathogenesis in cell or 
animal models of HD. Out of these sorted  $154$ HTT interacting proteins $17$ proteins are seen to interact with wHTT, $12$ proteins are identified in the aggregates of mutant HTT 
and $25$ proteins interact preferentially with mutant HTT. Thus in totality, wHTT interact with $17$ proteins whereas mHTT interacts with completely new $37$ proteins. The list of these protein  along with  other  subsidiary informations are presented in {\clrb Dataset S1 (sheet 1).}

\begin{figure}[h]
 \centering
\includegraphics[width=4cm,bb=14 14 475 349]{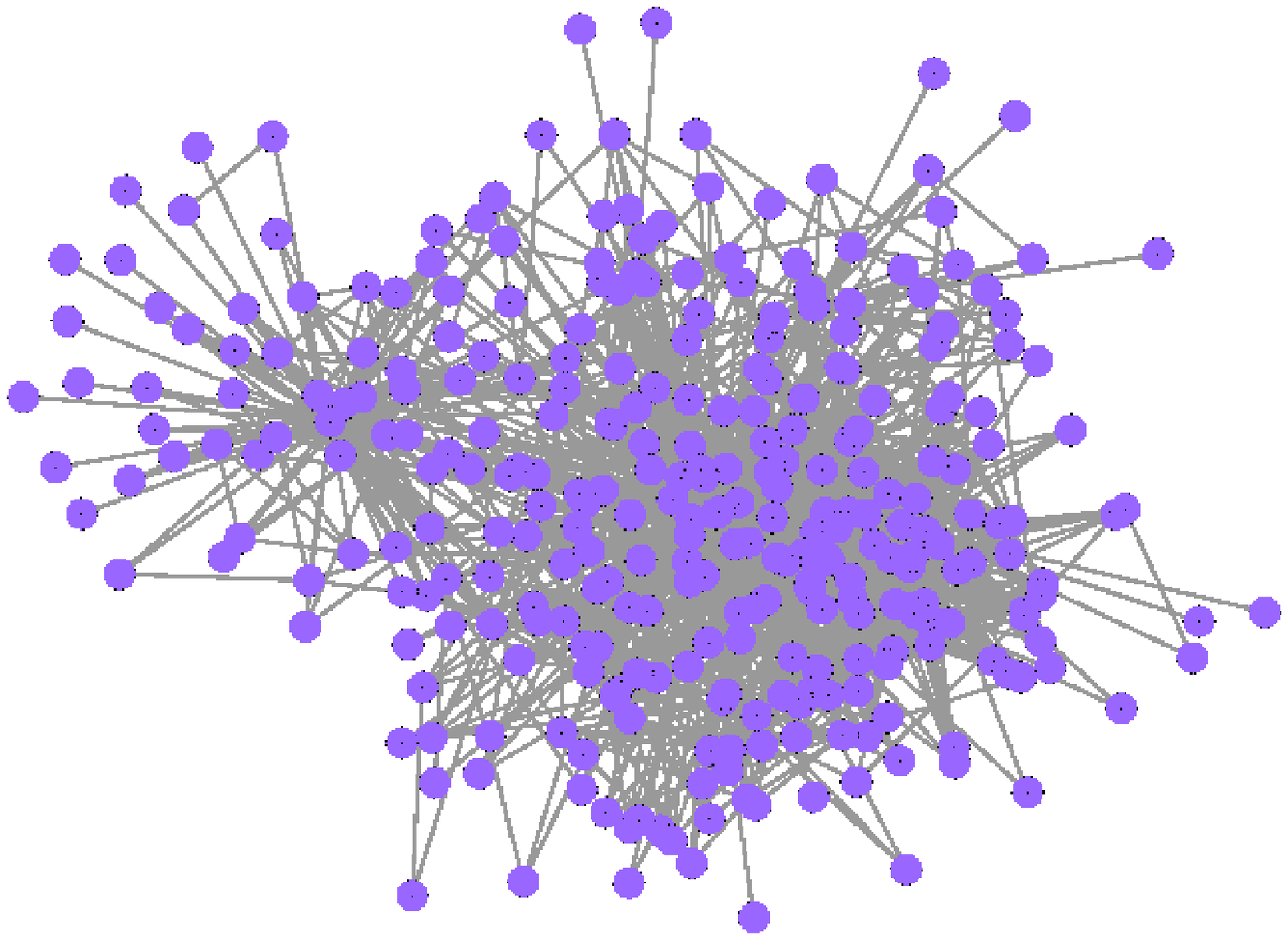}
\includegraphics[width=4cm,bb=14 14 370 348]{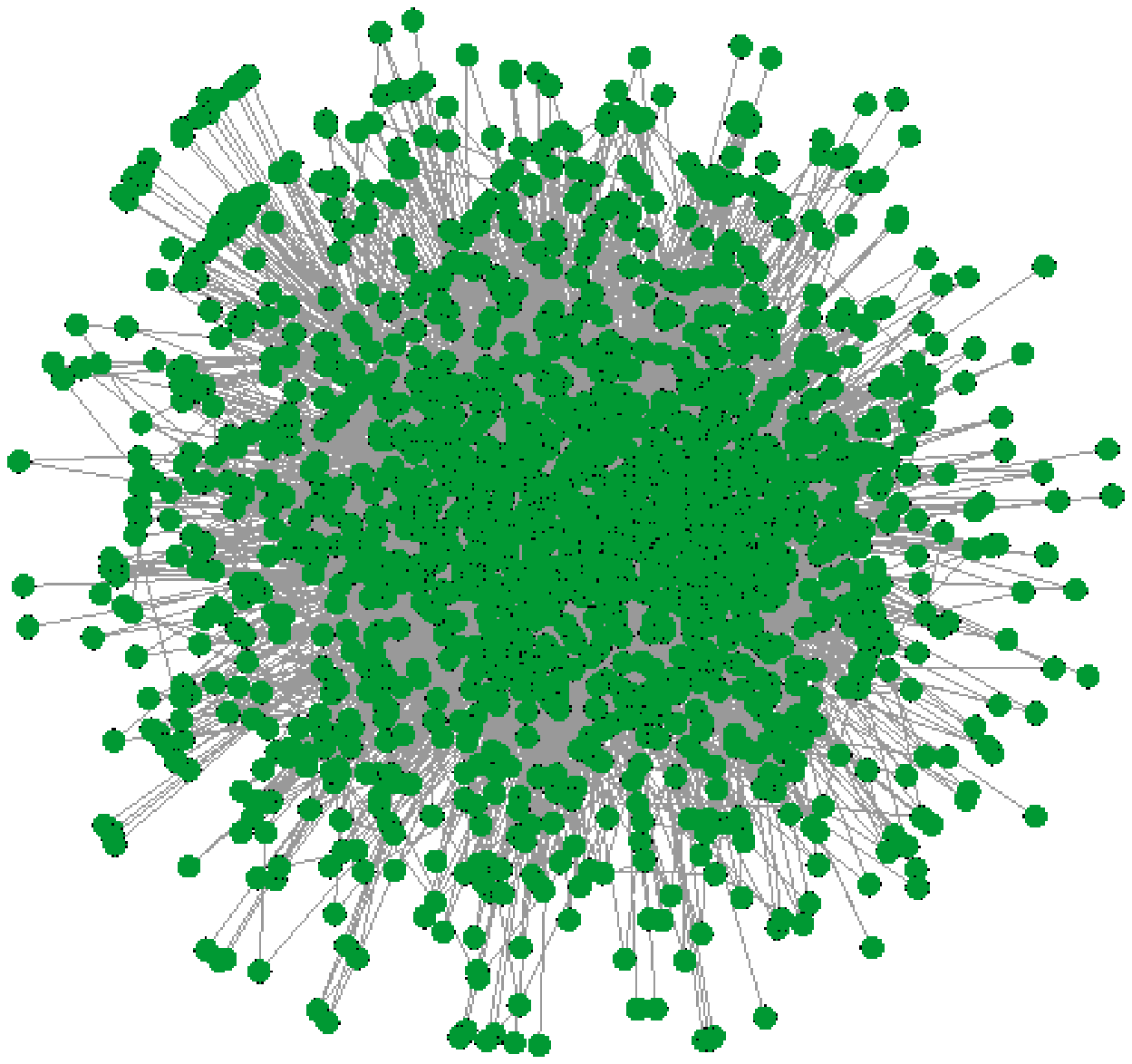}
\caption{ The interactome of wHTT and mHTT involves $306$ and $1542$ proteins respectively.
If a pair of proteins present in a particular set has interaction between them, then a link is constructed 
between these two proteins(nodes). Exhausting every pair of proteins for both the sets we end up in constructing
the interactome of the wHTT (left) and mHTT (right) using Cytoscape software \cite{Shannon}.}
 \label{fig:HTT_net}
\end{figure}

   The PPIN of  wHTT is constructed by taking   $306$  proteins (wHTT, its $17$ primary and  
$288$ secondary interactors)  and interactions among  them, whereas the  PPIN of 
mHTT  has $ 1542$ proteins ({\clrb see Dataset S1 (sheet 2 and sheet 3)}). Corresponding networks, drawn using  Cytoscape \cite{Shannon}, are  shown in {\clrb Fig. S\ref{fig:HTT_net}}.  

\subsubsection*{Common proteins in wHTT and mHTT networks}
Out of the $17$ primary interactors of wHTT, $8$ proteins namely (DLG4, CASP3, CTBP1, EIF2C2,
 REST, ZDHHC17, MEF2D, HIP1) change their direct interaction  
with wHTT and get involved in mHTT PPIN as secondary interactors. Whereas out of $288$ secondary interactors in wHTT PPIN, $10$ proteins now directly interact with mHTT. The list of these proteins are (PARK2, CASP10, TP53, CREBBP, SIN3A, CASP2, NFYC, CASP8, SP1, TBP). Also it is to be noted that, there are $107$ proteins remain as secondary interactors in both the wHTT and mHTT PPIN (list given in {\clrb Table S1}).

\subsection*{TP53  protein}
TP53 is a tumor suppressor gene. The mutation of this gene causes cancer. In a recent work   
Coffill $et.~al.$ \cite{CoffillCR} have studied the biological effects resulting from the missense mutations in TP53 (p53R273H). Their  experimental results indicate  that  wTP53 and mTP53 differentially interact with $17$ and $30$ proteins; list of these proteins are given {\clrb Dataset S5 (sheet 1).}

\begin{figure}[h]
 \centering
 \includegraphics[width=4cm,bb=14 14 386 360]{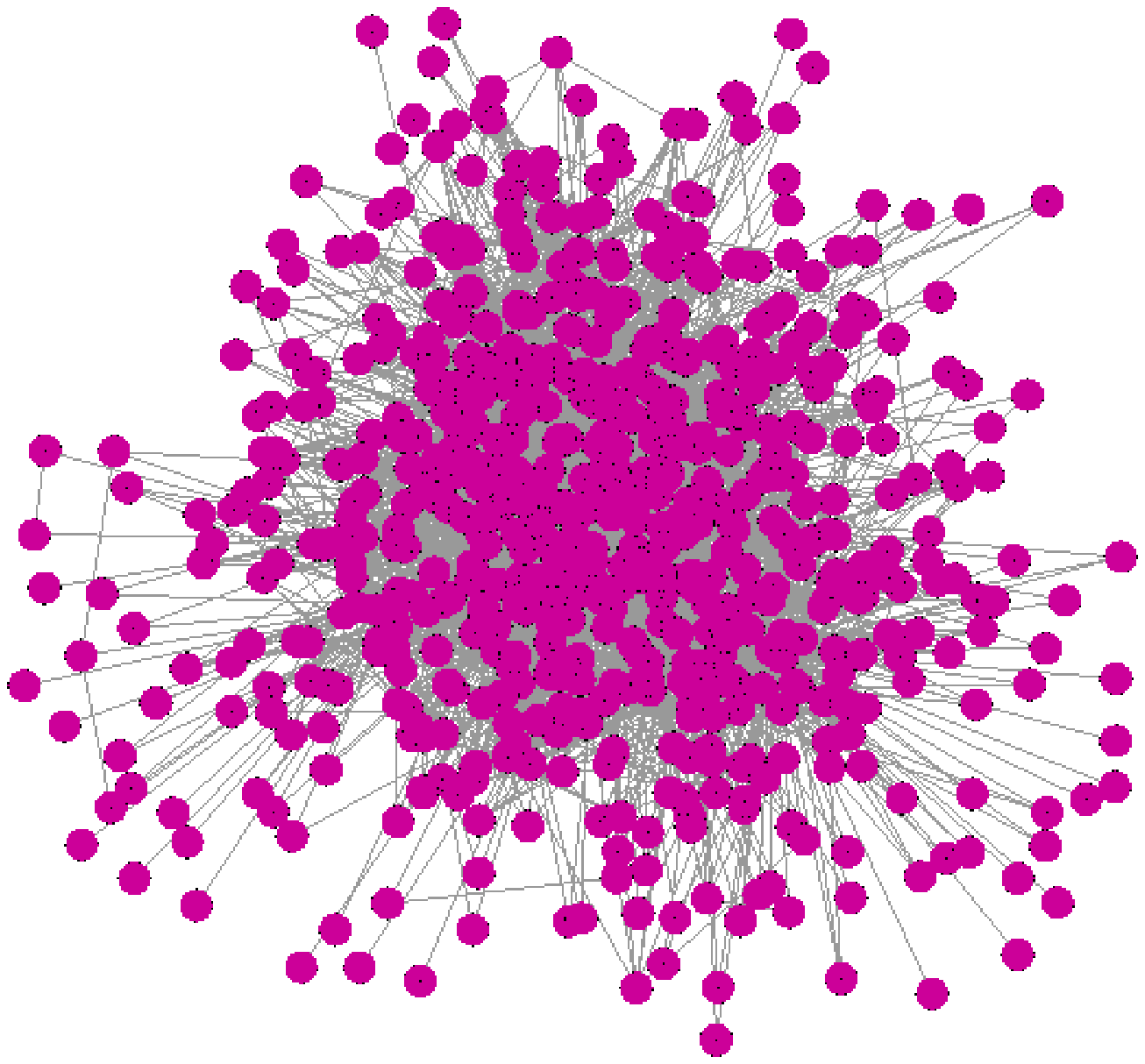}
\includegraphics[width=4cm,bb=14 14 385 356]{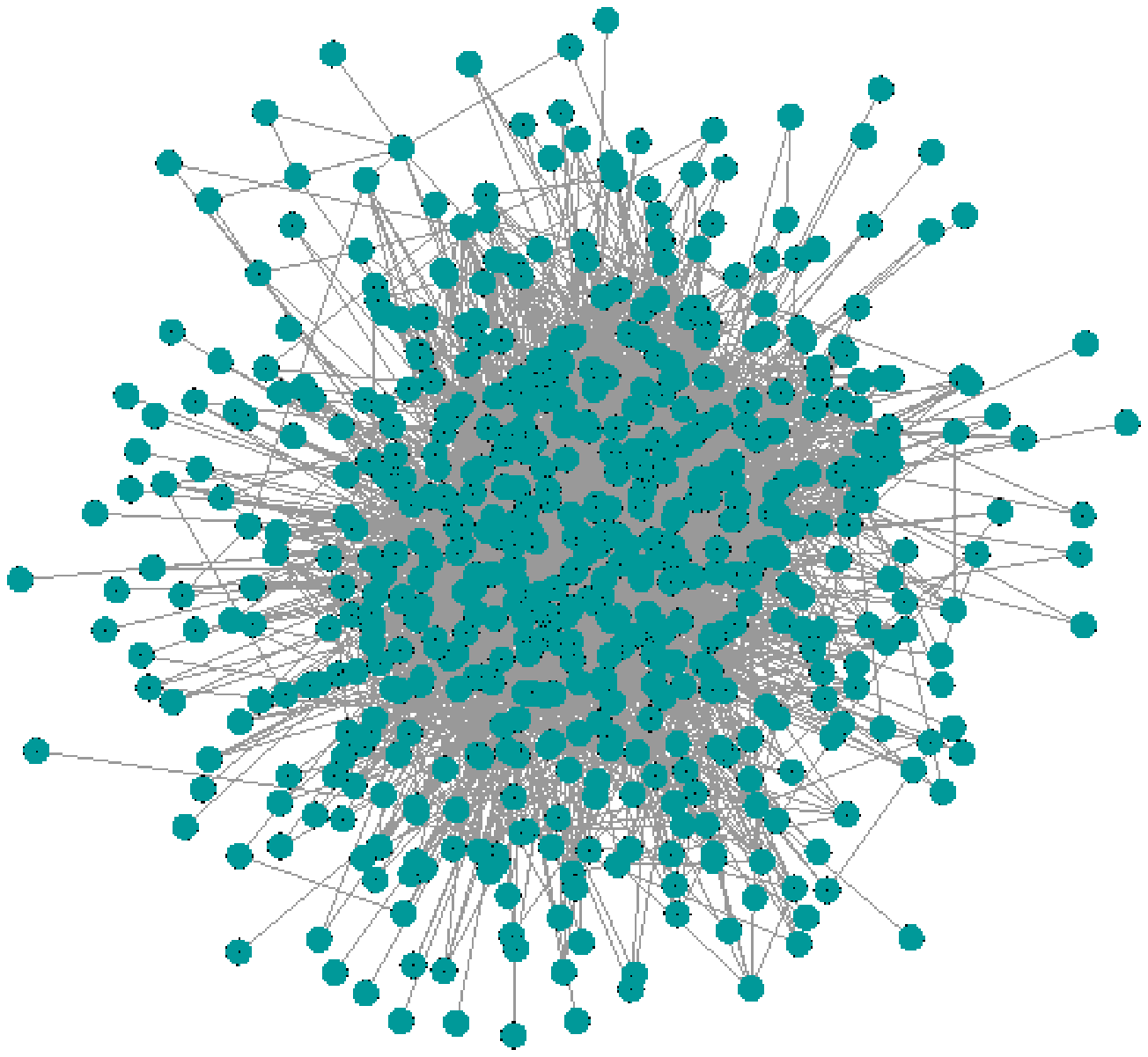}
 \caption{The PPIN of wTP53 and mTP53 interacting proteins are constructed using
the Cytoscape software \cite{Shannon}. The network in the left is for the wild TP53 and in right 
is for mutant TP53.}
 \label{fig:TP53_net}
\end{figure}

   The PPIN of  wTP53 is constructed by taking   $619$  proteins (wTP53, its $17$ primary and  
$601$ secondary interactors)  and interactions among  them, whereas the  PPIN of 
mTP53 has $578$ proteins ({\clrb see Dataset S5 (sheet 2 and sheet 3)}). The corresponding networks, drawn using  
Cytoscape \cite{Shannon}, are shown in {\clrb Fig. S\ref{fig:TP53_net}}.

\subsubsection*{Common proteins in wTP53 and mTP53 networks}
Similar to the previous case the mutation of TP53 causes change in interacting partners of TP53,
 altogether resulting in different secondary interactors.
Wild TP53 primarily interact with $17$ proteins, among them $5$ proteins 
(YWHAG, BAG6, TUBB3, RPL5, MDM2) lose their interaction with wTP53 and get involved in mTP53 PPIN as 
secondary interactor. Again these $7$ proteins (PARD3, PLK1, KHDRBS1, CSNK2A1, PSMD2, CSNK2B, PSMC1) 
which were secondary interactors in wTP53 PPIN now interact with mTP53 directly. Apart from these 
there are $111$ proteins common to both wild and mutant TP53 network, and are secondary interactors 
for both the networks (listed in {\clrb Table S1}).

\section*{Text 2 : Analysis of network structure }
In recent years  theory of complex networks  has been used enormously in  various fields \cite{Albert}.
 The social systems (friendship, mobile  or citation networks), information systems (World Wide Web and  
router networks), biological systems (neural, protein folding and gene regulatory networks) are a few to mention.  Network theory provides a quantifiable description for  comparison and characterization of complex networks. 
Here we shall focus on  some  robust measures of network topology :  the degree distribution, the clustering coefficient,  the  average path length and the diameter.

\textbf{ Degree Distribution}: The most elementary characteristic of a node (vertex) is its degree. The 
degree of a vertex $k$ in a network is defined as the number of edges connected to that vertex. 
The degree distribution, $P(k)$, gives the probability that a selected node has exactly $k$ links. 
$P(k)$ is obtained by counting the number of nodes with $k = 1,2, \dots$ links and  then normalizing  
by the number of nodes $N.$ 
The degree distribution  of many  real world  networks, including PPINs \cite{Reka} are 
commonly scale free, $i.e.$ $$P(k) \sim k^{-\gamma}.$$ 

\begin{figure}[h]
 \centering
 \includegraphics[width=8.5cm]{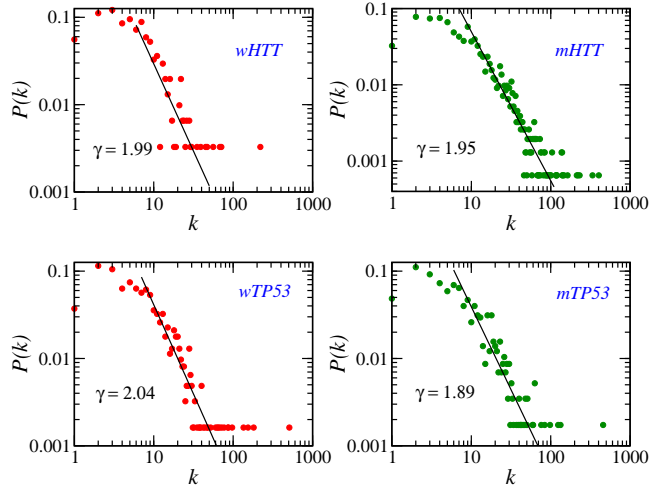}
 \caption{Degree distribution of wild and mutant HTT and TP53 network. All the four network follows scale free degree distribution. The corresponding power-law exponent is written in the graph itself.\label{fig:deg_dist}}
 \end{figure}

The degree distribution $P(k)$  for all  four  PPINs,  $i.e.$ 
wild and mutant networks of HTT and TP53,  are    found to be scale free (see {\clrb Fig. S\ref{fig:deg_dist}}),  
with  the exponent $\gamma \simeq 2.$   The average degree  and our best estimate of the exponent 
$\gamma$   are listed  in {\clrb Table S2}. 

\textbf{Clustering coefficient}: In a highly dense network, the neighbours of a given node
are  very likely to be connected among themselves. 
The clustering coefficient $$c_i = \frac{2E_i}{k_i(k_i-1)}$$
gives a  measure  of how connected  are the neighbours of node $i$.
Here $E_i$ the actual number of links  which connects all the $k_i$ neighbours of node $i$    
and $k_i (k_i-1)/2$  is the maximum number  of possible links.  Clearly  the maximum value of 
$c_i$ is unity.  The average clustering coefficient   of a network  of $N$ nodes is 
$$C=\frac{1}{N}\sum_i^N c_i,$$
which  gives a notion of how dense is the network, with  maximum value $1$  corresponding to a 
clique ($i.e.$ all $N$ nodes are connected to each other).

\textbf{Average path length :}  In  a connected network, a given node $i$ can be reached from any other 
node $j$ by  several  alternative paths. The shortest path  $l(i,j)$  is  the path with 
the smallest number of links between nodes $i$ and $j.$
The average path length   of a network  of $N$ nodes, thus $N(N-1)/2$  vertex pairs,  is   
$$L = \frac{\sum_{i \neq i}^N 2 l(i,j)}{N(N-1)}.$$

{\bf The network diameter :} The network diameter $D$ is the  largest  of  the shortest paths  among  all the vertex pairs. In other words,   $D$ is the length of the `longest' shortest path in the network. 
The  average path length and diameter for the networks of wHTT , mHTT, wTP53 and  mTP53  are calculated and their values are listed in the {\clrb Table S2}.

\textbf{ Modular structure of network :} Other than these network parameters, detection of modules or  community 
structures  is widely used to reveal  the organizational principles of nodes within the context of the entire system. Modules are   specific groups  of nodes   which has larger density of links  within  the group than  outside it. 
Each of such modules can be regarded as a discrete entity whose function or properties are in some way separable from other modules. A large variety of community detection techniques \cite{Fortunato} have been developed based on
centrality measures, link density, percolation theory etc. 

The most obvious way of finding  groups  in a network is to minimize the number of edges
connecting the groups. Recently, Newman $et. al.$ \cite{Girvan} proposed
a method of finding community structure of a network  by associating  a score called modularity $Q$
for each possible partition of a network;  a good partition of a network can be obtained by  
maximizing $Q.$  It is  argued that simply rearranging the network, such that only few edges 
exist between the communities, is not enough. Rather one must rearrange it in a way that the 
communities are connected with fewer than expected edges. 
Based on this,  Newman $et.$ $al.$   defined  a  score called modularity $Q$  which is proportional to the number
of edges present within the groups minus the expected number in an equivalent random network.

To be more precise,  let us consider an undirected network  of $N$  nodes labeled by $i=1,\dots N,$   
having an   adjacency  matrix $A$ with  boolean elements,  
\begin{equation*}
 A_{ij}=\begin{cases} 1, & \mbox{if  nodes }  i \mbox{ and } j \mbox{ are connected } \\ 0, &  \mbox{otherwise.} \end{cases}
 \end{equation*}
Thus, the degree of  the node  $i$ is $k_i= \sum_j A_{ij}$,  and the total  number of links  $m={1\over2} \sum_i k_i$ $={1\over2} \sum_{ij} A_{i,j}$ . If the  network  is to be partitioned into groups, the  modularity is  defined  by
 \begin{equation}
Q={1\over 2m}\sum_{ij}\left(A_{ij}- {k_i k_j\over 2m}  \right) \kappa_{i,j},
\label{eq:mdl}
\end{equation}
where  $\kappa_{i,j}=1$ (or $0$)   if  nodes $i$ and $j$  belong   to same (different)  partitions. 
Note that $k_ik_j/2m$ is the expected number of links  between nodes $i$ 
and $j,$ if edges were placed at random and it assures that $Q$ is maximum when two 
groups are  connected  by  smaller than expected number of links.    
A positive values of $Q$ indicates  the possible presence of community structure(s) and one 
need to look for a partition for which the modularity is preferably large and positive.
In the following we apply this procedure to obtain  modular structure of wHTT, mHTT, wTP53 and mTP53 networks. 


Modularizing the wHTT and mHTT network we see that wHTT network contains $7$ and mHTT network contains $8$ modules. Also the wTP53 and mTP53 networks divides into $4$ and $5$ modules.
The size of each modules $i.e.$ number proteins present in each modules are given in the {\clrb Table S2}. The names of the proteins that constitute the modules of wild and mutant HTT network are provided in the 
{\clrb Dataset S2 (sheet 1) and S3 (sheet 1)} respectively. Similarly for TP53, the name of the proteins present in the modules of the wild and mutant TP53 networks are given in the {\clrb Dataset S6 (sheet 1) and S7 (sheet 1)} respectively.

\section*{Text 3 :  Enriched biological processes in modules}
 To find the possible function(s) of the modules, we 
utilized bioinformatics enrichment tools. There are many tools available for such enrichment analysis, 
we mainly used GeneCodis3 \cite{Huang}  for the convenience.

GeneCodis is a web-based tool for interpretation of large scale data integrating from different sources like Gene Ontology (GO, pathways in Kyoto Encyclopedia of Genes and Genomes (KEGG), PANTHER pathways, Online Mendelian  inheritance in Man and others). 
 Given a set of genes in the query field, GeneCodis associates GO terms for each gene and calculate 
whether the fraction of genes in a particular GO term among the input gene list is over represented 
compared to the background frequency (the  total number of genes involved in  that GO term  over the 
total number gene in the organism). It further corrects the significance level for multiple testing. 
When  the list of proteins present in a given module   is  feed to GeneCodis3 as the query set, 
it gives a list of associated biological processes (GO terms)  and  their respective   
significance values. In this study we considered only those GO terms which have $p < 0.0001.$

\textbf{HTT PPIN :} The set of proteins present in each of the $7$ and $8$ modules 
of wHTT and mHTT PPINs are analyzed with GeneCodis3. The GO terms associated with 
each of the modules are enlisted in {\clrb Dataset S2 and S3} according to high to low significance. The GO terms 
with corrected significance value greater than $0.0001$ are discarded. Thus the number
 of GO terms belonging to each 
of the modules of wHTT PPIN are $W1(7),~W2(24),~W3(47),~W4(7),~W5(32),~W6(13),~W7(3)$
 and  for  mHTT PPIN are
$M1(161),~M3(78),~M5(198),~M6(12),$ $~M7(5).$ For the modules $M2,~M4$ and $M8$, all the GO terms 
that are obtained after GeneCodis analysis have $p$-value greater than $0.0001$, so these GO 
terms are not considered for further analysis.  Accumulating all the GO terms involved in the 
modules of wHTT and mHTT PPIN, one gets $129$ and $389$ GO terms respectively. (Note that merely 
adding the number of GO terms for each modules separately for wHTT and mHTT networks gives numbers 
greater than that mentioned; reason for this is that a single GO term can occur simultaneously
in two or more modules of wHTT or mHTT network.) A detailed study reveals that both these set 
of GO terms obtained for wHTT and mHTT modules has $65$ GO terms in common. Thus one can conclude 
that the unique GO terms that are involved only in wHTT and mHTT network are $64(=129-65)$ and
$324(=389-65)$. The Biological processes relating to these $64$ GO terms in wHTT network are lost
due to mutation-LOF, whereas the BPs corresponding to $324$ GO terms are GOF, and BPs for the
common ones $65$ are GOF/LOF both. We group the similar GO terms under a single broad BP.
A clear data demonstrating these LOF, GOF and GOF/LOF (see {\clrb Table S3}) is given {\clrb Dataset S4}.

\textbf{TP53 PPIN :} A similar study for the module of wild  and mutant TP53 PPIN modules is being 
done to obtain the LOF, GOF and GOF/LOF separately. The total  number of GO terms (considering GO 
terms with $p<0.0001$ for all the modules of wTP53 and mTP53 separately) that are present for 
wild and mutant TP53 networks are $127$ and $172$ respectively. These two sets has $70$ GO terms in 
common. Thus the unique GO terms present for the wild and mutant TP53 are $57$ and $102$; these are 
responsible for LOF and GOF function respectively. For the details of the loss and gain of functions 
refer to {\clrb Table S4} and {\clrb Dataset S8}.

\section*{Text 4 : Enrichment of metastasis  from GeneDeck }
Metastasis, which is shown as GOF  in mutation of  TP53\cite{CoffillCR},  is not described for ``biological process'' in Gene Ontology term. We have used another tool, GeneDecks \cite{Safran},
which provides a similarity metric by highlighting shared descriptors between genes, based on annotation within the GeneCards compendium of human genes.  Given a set of genes as query to the field “Set Distiller” in GeneDecks (Version 3), an online analysis tool that provides output of various descriptors like various cancers and diseases with  the attribute “Disorder” taking data from different resources like Alma Knowledge Server, PharmGKB, The Breast Cancer Gene Database, Tumor Gene Database, GeneTests (formerly GeneClinics), OMIM, Swiss-Prot and Genatlas. 
The proteins present in the modules of wTP53 and mTP53 are analyzed using GeneDeck in order to get the information about the presence of metastasis in the corresponding modules. It is seen that the almost all the modules of wTP53 (except module $W4$) and mTP53 shows enriched metastasis (refer to  {\clrb Dataset S9 }).

\newpage

\begin{table}[h]
\begin{center}
\tiny{
\caption{List of the secondary interacting proteins that are common in to both wild and mutant
 PPIN of HTT and TP53 separately.}
    \begin{tabular}{l| p{10cm} }
Common  & Protein name  \\
\hline
wHTT-mHTT($107$) & YWHAQ, NFATC2, BCL3, HIST3H3, SMARCA4, TBK1, MAPK1, BTRC, SMARCC2, APC, PRKCA, CTTN, SMARCE1, UBC, MEN1, WAS, TCF7L2, KDM1A, NEUROD1, CASP6, RBBP5, GTF2E2, CBL, ZEB1, NUP98, AKT1, NUB1, IFT57, DHX58, BLM, TUBG1, EHMT1, COPS5, EHMT2, KDM5C, MAPK14, PSMC3, CFLAR, CARM1, NCOR1, CASK, MDM2, PRMT5, NCOA3, HSPB1, CDC42, BRCA1, HTR2C, HSPD1, FZD7, TBL1X, UBE2I, SENP3, HDAC1, HDAC2, HEY2, HDAC3, HDAC4, SUB1, SREBF2, HDAC5, TP73, THAP11, EP300, BID, HDAC9, TCF4, GTF2B, WIPF1, EGFR, KAT2A, SNCA, KAT2B, MLL, PFDN1, GRIK5, SERPINH1, RPS10, DNM1, SUMO1, SUMO2, USO1, KAT8, CHD3, RAC1, SIN3B, IKZF1, HIC1, DBNL, GRIN2B, IKZF4, SOX2, EIF6, YWHAE, DICER1, MAP3K14, BIRC2, MECOM, TGIF1, KHDRBS1, BIRC5, CABIN1, SATB1, HGS, ABL1, BCL2, XIAP \cr
\hline
wTP53-mTP53($111$) & UBC, TANK, RBBP4, RBBP7, PPP4C, MAGED1, SETDB1, SMAD1, UIMC1, CDH1, TERF1, NOLC1, PSMD10, CCDC85B, UBE2I, USP42, IKBKB, WRAP73, DCAF7, E2F1, MGMT, NCL, HSP90AA1, PRMT1, SRRM1, SRRM2, CHEK2, FBXW11, ITCH, PIK3R1, RB1, KIF5C, YWHAZ, STUB1, ARRB1, EP300, AR, SNAI1, RPS6KB1, MAP3K5, ESR1, UCHL1, UCHL5, STAU1, PCK1, MDM4, RPA1, RPA2, RPAP1, BCL3, CDC5L, RASSF8, HSPA1A, PSMA3, NCOR2, PSMA7, TP53BP1, NDRG1, EXO1, CASP8, ARAF, H2AFX, PRPF40A, PELP1, EIF2C3, PIN1, NR3C1, CLSPN, RAD21, NEDD4, NEDD8, IMMT, CHAF1A, TSC1, MTA1, PUF60, CTBP1, MLLT4, MAPK6, TNFRSF1A, MEPCE, ARR3, SRGAP2, CREBBP, SRC, PRKCZ, SFN, CDK9, ATM, WNK1, RIPK1, SUMO1, SUMO2, HDGF, BTRC, LRIF1, CDKN1A, MDC1, RNPS1, PPP2R2D, HIST1H3A, HDAC1, HDAC3, DISC1, HDAC5, HDAC6, TP73, TAF1, PSMC4, PSMC5, C5ORF25\cr
\hline
  \end{tabular}
}
\end{center}
\label{tab:table1}
 \end{table}

\begin{table}[h]
 \begin{center}
\small{
\caption{Comparison of network properties between wild-type and mutant PPIN of HTT and TP53.}
\label{table:net-prop}
\begin{tabular}{lrrcccclll}\hline
 network &node & edge \footnote{The total number of edges is written as the sum of the secondary links and primary links.} & $\langle k \rangle $ &$\gamma$ & $C$ & $d$ & $L$  & No. of Modules: Module sizes & Q \cr \hline
wHTT & 306 & 1380+17 & 9.13 & 1.95 & 0.436 & 4 & 2.418  & 7: 18, 66, 79, 18, 82, 8, 35 & 0.415\cr 
mHTT & 1542 & 13105+37 & 17.05 & 1.99 & 0.361 & 4 & 2.349  &8: 643, 3, 377, 2, 485, 7, 22, 3 & 0.302\cr   
wTP53 & 619 & 3718+17  & 12.07 & 2.04 & 0.452 & 4 & 2.232  &4: 204, 151, 183, 81 &  0.331 \cr 
mTP53 & 578 & 3521+30  & 12.29 & 1.89 & 0.406 & 4 & 2.282  &5: 193, 127, 25, 111, 122 & 0.338\cr 
\hline 
\end{tabular}
}
\end{center}
\end{table}




\begin{sidewaystable}
\begin{center}
\caption{LOF, GOF, LOF/GOF occurred due to mutation of HTT.}
\begin{tabular}{p{6.5cm}lll}
\hline
GO terms related to & Modules of mHTT network & Modules of wHTT network & Modules of wHTT-mHTT network \cr
 &(unique GO terms) & (unique GO terms) & (unique GO terms)\cr
\hline
Cell cycle & M1(4),M3(12),M5(5) &  W2(1),W6(3) & (W2,W5)-(M1,M3,M5)(4) \cr
Signaling & M1(30),M5(31),M6(2) & W3(17),W4(4) & (W3,W4,W5,W7)-(M1,M5)(9) \cr
Transcription processes and regulation & M1(5), M3(5), M5(31) & W7(1) & (W1,W5,W7)-(M3,M5,M6(17) \cr
Apoptosis & M1(11) & W2(1) & (W2,W5)-(M1,M3,M5)(12) \cr
DNA replication & M3(12) & - & - \cr
DNA damage and repair & M1(6),M3(17),M5(4) & - & (W3,W5)-(M1,M3,M5)(4)  \cr
Immunological & M1(7) & - & - \cr
Protein folding & M1(7) & - & - \cr
Autophagy & M1(5) & - & - \cr
Translation & M1(3) & - & - \cr
Metabolism & M1(1) & - & - \cr
Development and differentiation & M1(4),M5(57),M6(8) & - & (W3,W5)-M5(5) \cr
Cell migration and shape & M1(4) & - & - \cr
Proteasomal degradation & M1(14) & W2(1) & (W2,W5)-(M1,M3)(3) \cr
Carbohydrate/Glucose transport/metabolism  & M5(4) & W6(5) & -\cr
Protein complex/membrane assembly/stabilization & M1(9) & - & -  \cr
Cell growth  & M5(7) & - & - \cr
Transcription from RNA polymerase III & M7(4) & - & - \cr
Others & M1(4),M3(5),M5(6) & W3(4) & (W2,W3,W4,W5,W6)-(M1,M5)(6) \cr
Gene silencing/micro RNA and RNA processing /translation & - & W1(5) & - \cr
Synaptic transmission and neuronal activity & - & W3(12) & W3-M1(2) \cr
Transport (ion/sugar) & - & W3(5) &  -\cr
Circadian rhythm & - & W2(1) & - \cr
Protein /transmembrane  transport & - & W6(4) & - \cr
NFKB1 regulation  & - &-  & W2-M1(2) \cr
Proliferation/growth  & - & - & W5-(M1,M3,M5)(1) \cr
\hline
\end{tabular}
\end{center}
\end{sidewaystable}


\begin{sidewaystable}
 \begin{center}
  \caption{LOF, GOF, LOF/GOF occurred due to mutation of TP53.}
 \begin{tabular}{p{5cm}lll}
\hline
GO terms related to & Modules of mTP53 network & Modules of wTP53 network & Modules of wTP53-mTP53 network \cr
 &(unique GO terms) & (unique GO terms) & (unique GO terms)\cr
\hline
Cell cycle & M1(3),M2(4),M5(1) & W3(3) & (W1,W3,W4)-(M1,M2,M5) (14)\cr
Signaling & M1(13), M5(7), M2(1) & W1(8), W3(1), W2(4) & (W1,W2,W3)-(M1,M4,M5) (9) \cr
Transcription processes and regulation & M5(6) & W3(13), W4(2), W7(1) & (W1,W2,W3,W4)-(M1,M2,M5) (15)\cr
Apoptosis & M5(1) & W2(2) & (W1,W2,W3,W4)-(M1,M2,M5) (4) \cr
DNA replication & M2(11) & W3(3) & W3-(M2,M5) (3) \cr
DNA damage and repair & M1(1), M2(14), M5(2) & W3(1) & (W1,W2,W3,W4)-(M1,M2,M5) (10)\cr
Immunological & M2(3) & W2(3) & (W2,W4)-(M1.M2) (2) \cr
Development and differentiation & M5(11) & - & W3-M5 (1) \cr
Proteasomal degradation & M1(2),M2(3) & W1(1), W4(2) & (W4)-(M1.M2) (3)\cr
Cell-cell communication & M1(5) & - & -\cr
Cell proliferation/growth & M5(4) & - & W3-M5 (2) \cr
Protein complex/membrane assembly/stabilization & M1(4) & - & -\cr
Others & M1(2), M5(4) & W1(1), W4(4) & (W1,W2,W3,W4)-(M1,M2,M5) (5)\cr
Translation & - & W1(1), W4(5) & -\cr
Cell Migration and movement  & - & W1(2) & -\cr
Circadian rhythm & - & W4(1) & -\cr
Metabolism & - & - & W4-M2 (2)\cr
\hline
\end{tabular}
 \end{center}
  \end{sidewaystable}